\documentclass[letterpaper]{jpconf}
\usepackage{graphicx}
\begin{document}

\title{Correlations in STAR: interferometry and event structure}

\author{Mikhail Kopytine for the STAR Collaboration}

\address{Department of Physics, Kent State University, Kent, Ohio, USA }

\ead{kopytin@bnl.gov}

\begin{abstract}
STAR observes a complex picture of RHIC collisions where correlation
effects of different origins -- initial state geometry, semi-hard
scattering, hadronization, as well as final state interactions such as
quantum intensity interference -- coexist. Presenting the
measurements of flow, mini-jet deformation, modified
hadronization, and the Hanbury Brown and Twiss effect, we
trace the history of the system from the initial to the final
state. The resulting picture is discussed in the context of
identifying the relevant degrees of freedom and the likely
equilibration mechanism.
\end{abstract}

\section{Introduction}

This talk is an overview of recent STAR results in the field
known as "correlations and fluctuations".
 Out of multiple reasons to
analyze correlations and fluctuations at RHIC known before the data
became available, there is one that seems to dominate now: by
measuring correlations and their evolution we learn about
{\it equilibration} of the system. Is it taking place? What is the
mechanism? And what is equilibrating? 
Novel and advanced data analysis techniques have been created lately
to address these issues.
I will focus on the developments since Quark Matter 2004.

Our evidence for equilibration includes elliptic flow, medium
modification of mini-jets and of charge-dependent correlations.
Related issues are the azimuthally-sensitive HBT and the blast wave
picture.

\section{Elliptic and directed flow as a particle number correlation}
\label{Sec:flow}
\begin{figure}[t]
\includegraphics[width=18pc]{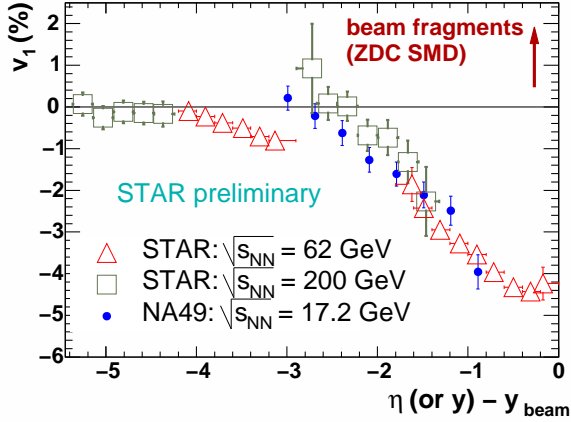}\hspace{2pc}%
\begin{minipage}[b]{18pc}\caption{\label{Fig:v1} 
Directed flow coefficient as a function of pseudorapidity relative
to the projectile beam in STAR and NA49}
\end{minipage}
\end{figure}
In non-central collisions the initial spatial configuration is anisotropic.
But the configuration space anisotropy does not produce momentum space
anisotropy unless there is another mechanism involved, such as
 local equilibration via rescattering.
If the latter is occuring, then it is sensible to talk about pressure gradients, and the azimuthal
variation thereof.
This picture naturally calls for analysis of correlations in velocity
between local elements of expanding matter, but experimentally one is limited to final
state particles.
The flow analysis in this field began 
with diagonalization of the ({\it momenta}-based) sphericity tensor 
~\cite{Plastic_Ball_review}.
In experiments with limited acceptance in rapidity at higher energies
(AGS, SPS and RHIC) a different technique~\cite{Voloshin_Zhang_Poskanzer} 
became popular which 
utilizes the observed azimuthal anisotropy by expanding the
particle {\it number} distribution with respect to the reaction plane angle $\Psi_r$ 
into a Fourier series
\begin{equation}
E\frac{\,d^3N}{\,d^3p}=\frac{1}{2\pi}\frac{\,d^2N}{p_t\,dp_t\,dy}
\{1+\sum_{m=1}^{\infty}2v_m \cos[m(\phi-\Psi_r)]\},
\label{Eq:v_m}
\end{equation} 
with $v_2$ typically being the largest component. 
Cumulant-based analyses, developed later~\cite{STAR_long_flow_paper}, 
distinguish between two-particle and higher order
correlations and separate flow (a multiparticle correlation) from
the lower-order "non-flow".

STAR's recent measurements of directed flow ($v_1$),
shown in Fig.\ref{Fig:v1}, use 3-particle
cumulants to minimize the "non-flow" effects.  In the limiting
fragmentation picture~\cite{limiting_fragmentation}, 
what matters for the fragmenting system is the
relative rapidity with the nearby spectators, hence the choice of the
abscissa. 
The data from different energies are consistent with limiting
fragmentation.  
The $v_1$ term requires knowledge of $\Psi_r$
between 0 and $2\pi$ and thus an azimuthal direction 
rather than orientation of the reaction plane. This direction, given by 
the spectators, is reconstructed event-by-event by means of a 
position-sensitive zero-degree shower-maximum detector (ZDC SMD) which 
selectively detects fragmentation neutrons behind a RHIC dipole magnet.
These measurements indicate that the directed flow at RHIC is opposite in azimuthal angle to
the direction of the spectators (an ``anti-flow'').
We will return to the velocity aspect of the elliptic flow in Sections \ref{Sec:HBT} and
\ref{Sec:pt_correlation}.

\section{Final state geometry}
\label{Sec:HBT}
\begin{figure}[t]
\begin{minipage}[t]{22pc} 
\includegraphics[width=24pc]{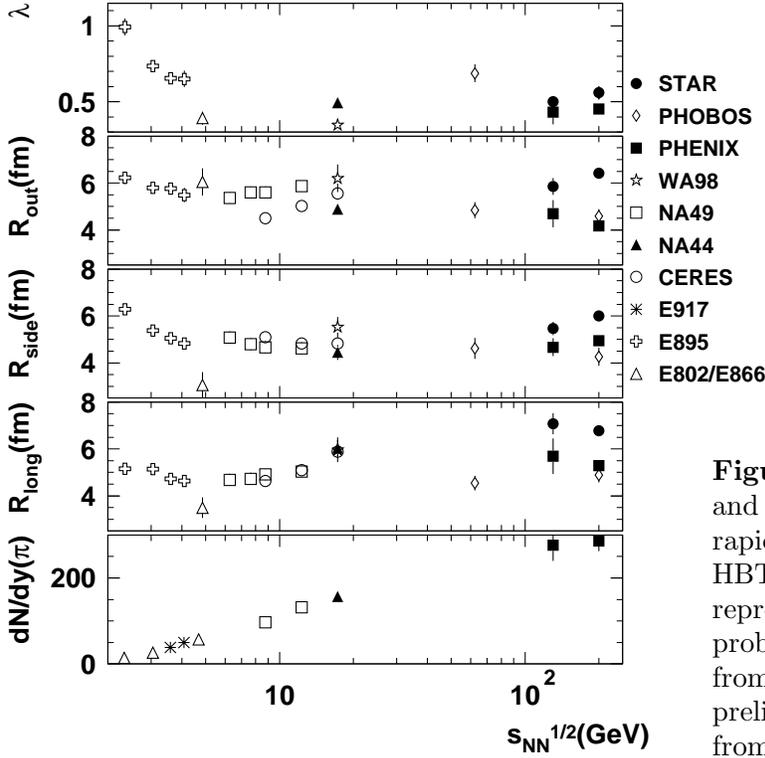}
\end{minipage}
\begin{minipage}[b]{15pc}
\caption{
\label{Fig:HBT_sqrt_s}
Fit parameters for $\pi^-\pi^-$ and $\,dN/\,dy$ of $\pi^+$ at or near mid-rapidity
as a function of $\sqrt{s_{NN}}$. The HBT data are either $p_t$ integrals or represent a $p_t$ bin
closest to the most probable inclusive $p_t$ and are taken from 
~\cite{HBT_list}.
The NA49 HBT data are preliminary.
The $\,dN/\,dy$ data are from 
~\cite{dNdy_list}.}
\end{minipage}
\end{figure}

While $v_m$ characterize the particle distribution in 
 momentum space, observation of the corresponding structure in the configuration space
 rests with the Hanbury Brown-Twiss (HBT)
 interferometry.   
Herein a correlation  function is defined as the ratio of two-particle probability density to
 the uncorrelated reference, which in STAR is obtained by forming mixed pairs. 
In the Cartesian  parameterization,
\begin{equation}
C_{\rm fit}(\vec{q}) = 1+
\lambda \exp(-q^2_o R^2_o -q_s^2R_s^2 - q_l^2R_l^2),
\label{Eq:Bertsch-Pratt}
\end{equation}
the momenta of a pair are projected onto the transverse plane
where one identifies directions of their sum and difference.
Projections of
the momentum difference on those directions become, respectively, outward $q_o$ 
and sideward $q_s$ components. 
The longitudinal component $q_l$
is orthogonal to both of them. Corresponding to $q$ are the radii parameters $R$.
Fig.\ref{Fig:HBT_sqrt_s} shows a $\sqrt{s_{NN}}$ dependence of the quantities
representing the best fits of the parameters in Eq.\ref{Eq:Bertsch-Pratt} to the experimental data.
The bottom panel presents density of $\pi^+$ multiplicity per unit of rapidity, 
which displays a steady
logarithmic increase with $\sqrt{s_{NN}}$.
The latter may be contrasted with the general lack of energy dependence of the radius parameters
which underscores the non-triviality of interpreting the HBT results.
In the hydrodynamical framework the concept of {\it length of homogeneity}\footnote{The 
term ``effective length'' was used in the original article.}
has been introduced~\cite{homogeneity_length}, and it is known that the measured radii reflect 
those lengths\footnote{This result was derived for locally thermally equilibrated systems.}
(or rather, according to ~\cite{Weiner_text}, a combined effect of the 
correlation lengths in the {\it primordial correlator} between
the {\it currents} emitting the pion radiation and the source size).
These lengths are smaller than the geometrical source sizes in an
expanding (due to pressure gradients) and thus inhomogeneous
system.  
A more subtle issue is that of the little relative
difference between $R_{\rm o}$ and $R_{\rm s}$, which is believed to
increase with longer emission.

The quantitative understanding of this information remains a
model-dependent problem.  In particular, one may
assume a local, thermally-equilibrated system and thus ascribe
all the sources of inhomogeneity (dynamical correlations) to
hydrodynamic expansion.  Taking this approach and fitting the data
(including single-particle spectra) with the modified Blast Wave
model, Retiere and Lisa obtained a set of
parameters~\cite{Blast_Wave_model}.  Remarkably, the
azimuthally-varying component $\rho_2$ in their description of the
transverse velocity field appears to be non-zero and well constrained
by the fit.  This represents an indirect measurement of the
transverse-velocity aspect of elliptic flow which complements the
particle-number-related $v_2$.

The azimuthally-dependent radii recently published by
STAR~\cite{STAR_asHBT} show an effective source elongated
perpendicularly to the reaction plane.  This indicates shorter
homogeneity lengths (and thus larger gradients) in-plane than
out-of-plane, which is consistent with the geometry of the early stage
of the collision.  The early emission is consistent with short
emission time.


\section{Charge-dependent number correlations-- modified hadronization in the medium}
\label{Sec:CD}
\begin{figure}
\includegraphics[width=4.9cm]{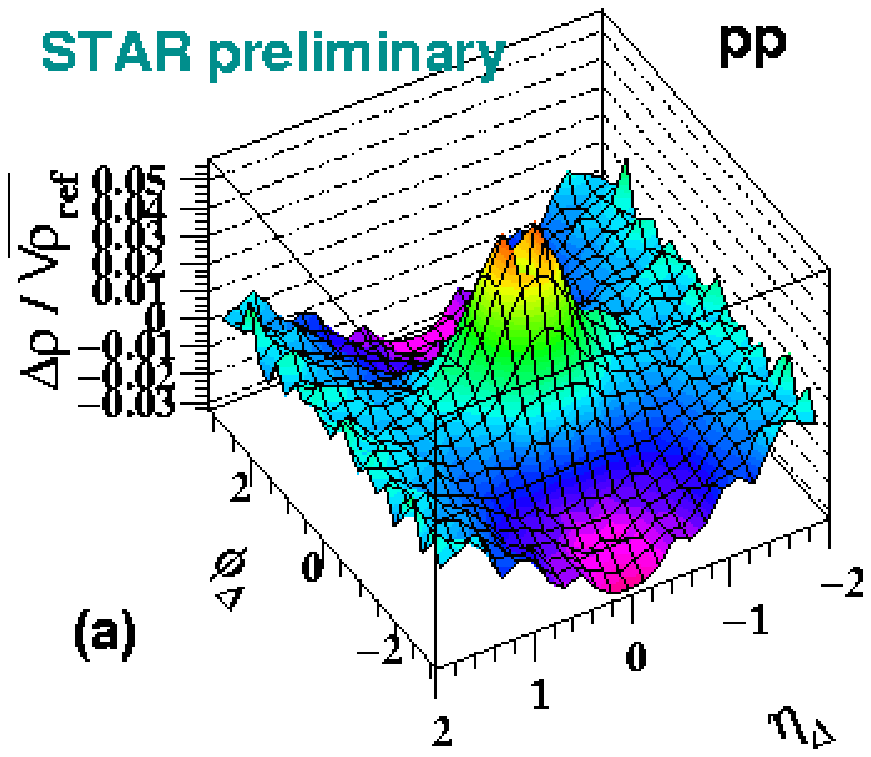}
\includegraphics[width=4.4cm]{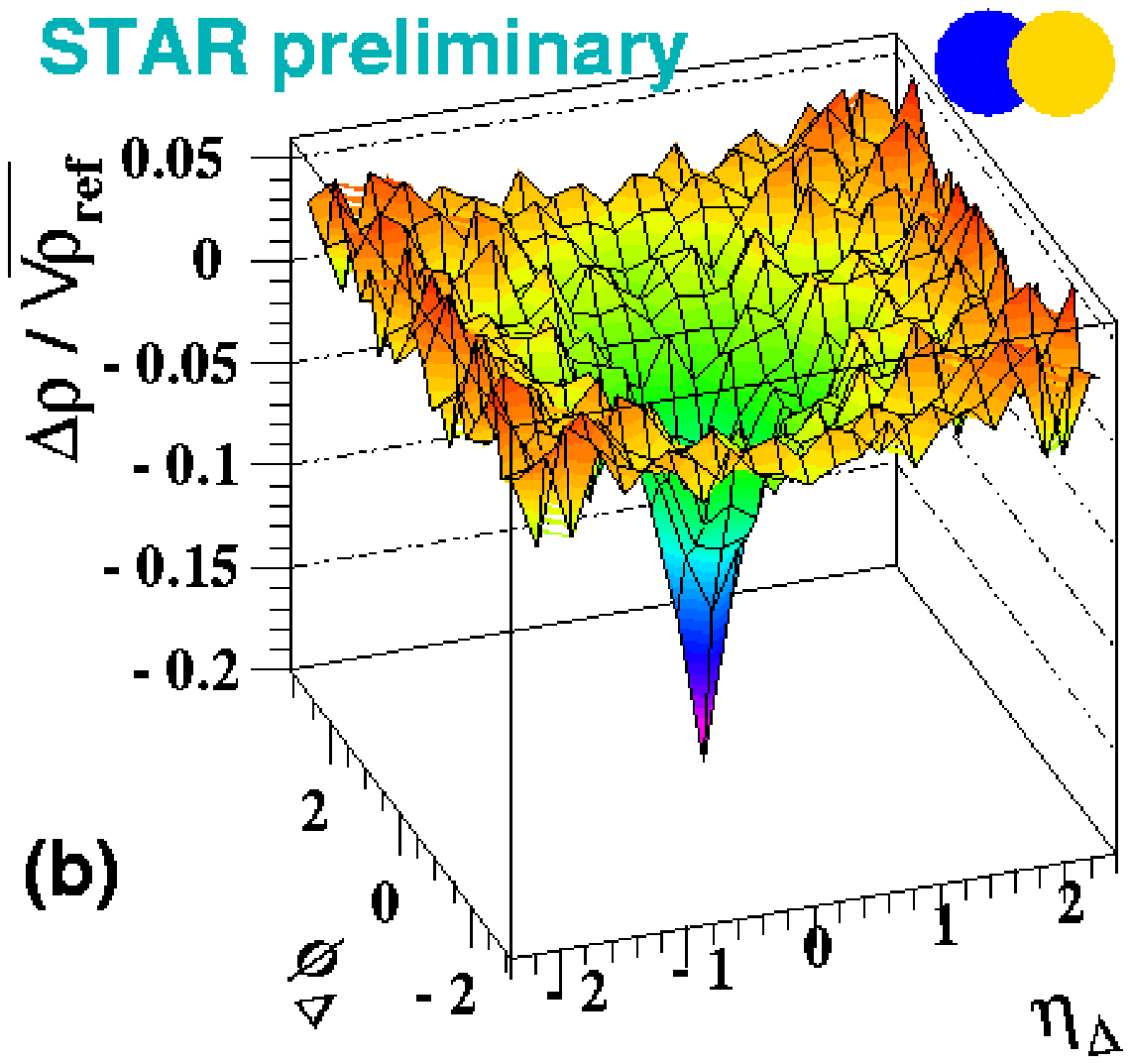} 
\includegraphics[width=4.4cm]{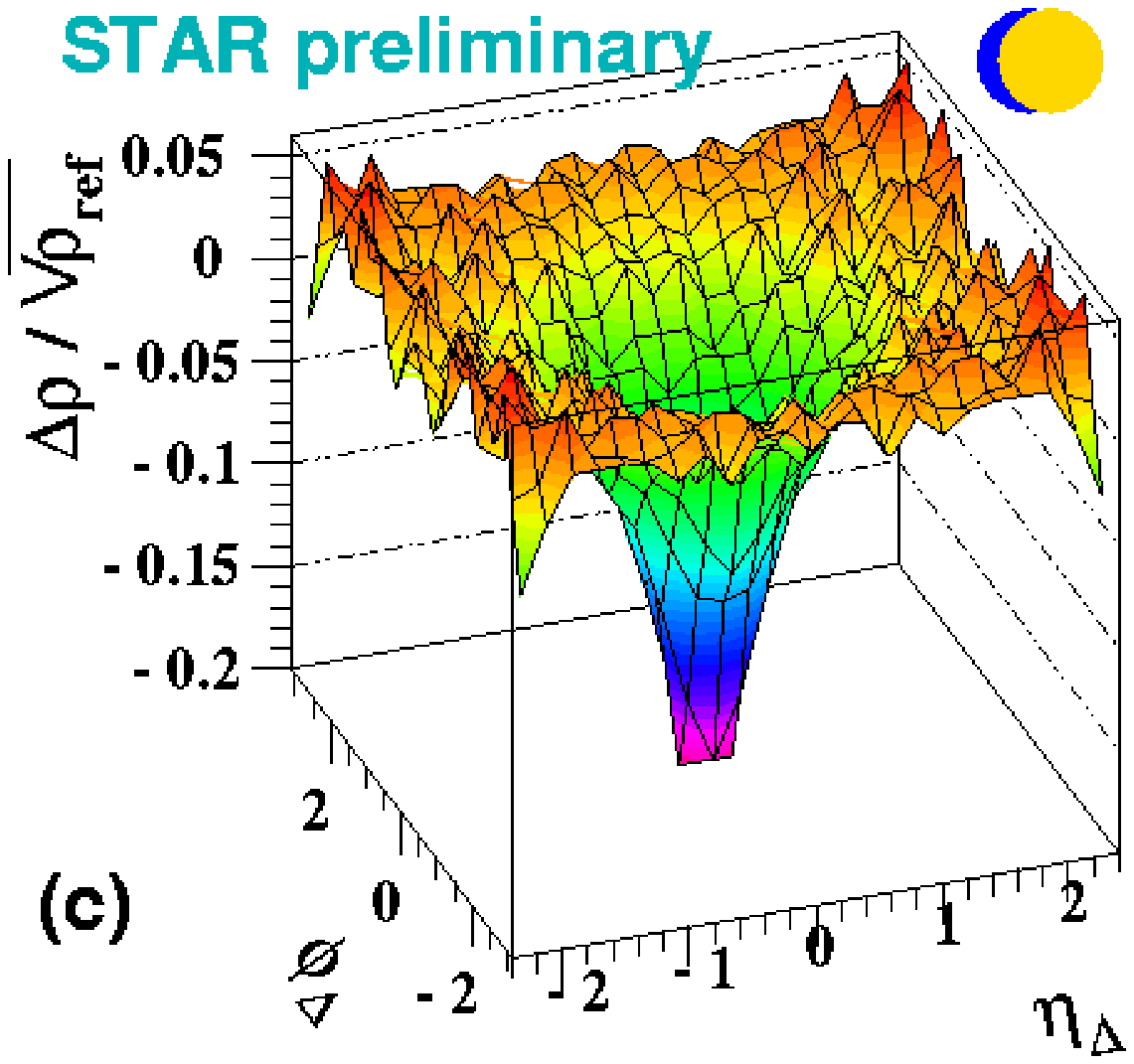}
\caption{
\label{Fig:CD}
Comparison of  CD correlations in $\eta$ and $\phi$: (a) pp
collisions at $\sqrt{s}=200$ GeV, (b) peripheral AuAu collisions, (c)
central AuAu collisions.  (b) and (c) are at $\sqrt{s_{NN}}=130$
GeV. 
}
\end{figure}

The charge-dependent (CD) number correlation is defined as a difference between the like-sign and
the unlike-sign correlations.
(In this and subsequent sections we discuss results obtained without particle identification).
The correlations are defined by comparing same-event pairs with mixed-event pairs, obtained from
mixing similar events.
To characterize and normalize correlations, we use
\begin{equation}
\frac{\Delta\rho}{\sqrt{\rho_{\rm ref}}} = \frac{\rho_{a,b}-\rho_a\rho_b}{\sqrt{\rho_a\rho_b}},
\label{Eq:Delta_rho}
\end{equation}
where $a$ and $b$ index particles (or in the discretized case, bins in the kinematic space occupied
by particles),
$\rho_{a,b}$ is pair density distribution, while
$\rho_a$ and $\rho_b$ are the corresponding single-particle distributions, whose product forms the
reference density $\rho_{\rm ref}$.
Experimentally one deals with fluctuating histogram bin contents $n_a$
and $n_b$ and Eq.\ref{Eq:Delta_rho} becomes (see \ref{Sec:Notation})
\begin{equation}
\frac{\Delta\rho}{\sqrt{\rho_{\rm ref}}} \rightarrow 
\frac{\overline{n_a n_b} - 
\overline{n_a}\times \overline{n_b}}{\sqrt{\overline{n_a}\times\overline{n_b}}}
=\frac{{\rm Cov}[n_a,n_b]}{\sqrt{\overline{n_a}\times\overline{n_b}}}
\end{equation}
If the variance in the bin population $n$ is dominated by the Poissonian process,
${\rm Var}[n] = \bar{n}$, and if in presence of correlations
${\rm Cov}[n_a,n_b] \propto {\rm Var}[n]$ (number of correlation sources is proportional
to multiplicity) this normalization eliminates the trivial multiplicity dependence of the
correlation amplitude.
Eq.\ref{Eq:Delta_rho} represents a {\it per-particle} measure of correlations.
\footnote{
In this terminology, the correlation function commonly used in HBT represents a 
{\it per-particle-pair}-normalized quantity.
The choice of normalization depends on the correlation mechanism under study.
}

In paper~\cite{STAR_CD}, CD correlations in $\eta$ and $\phi$ have been
analyzed simultaneously by using {\it joint autocorrelation}.
Fig.\ref{Fig:CD} shows a comparison of CD correlations in pp collisions (panel (a)) with
peripheral and central AuAu collisions.
Kinematic cuts on the track separation
are applied to eliminate the
contribution of HBT and Coulomb effects in AuAu collisions (panels (b) and (c))~\cite{STAR_CD}. 
A much broader HBT correlation for pp collisions contributes the
gaussian peak at the origin in (a). Transverse momentum conservation
also contributes to the structure at the origin (the ridge beneath the
gaussian peak). The main feature, the negative gaussian peak on
$\eta_\Delta$, is attributed to local charge conservation during
hadronization. 
If the process happens on a longitudinally
expanding string (as e.g. in Lund model), the strict alternation of 
positive and negative hadrons in rapidity holds (modulo interspersed neutrals).
This implies that $\eta_\Delta$ for an unlike-sign pair tends to be shorter than for the like sign,
regardless of $\phi_\Delta$, creating a groove in the CD plot, clearly seen in Fig.\ref{Fig:CD}(a).
The $\phi_\Delta$-independent groove appears to diminish in amplitude from (a) and (b) to (c).
In addition, the negative peak at 0 relative angle changes its shape, 
becoming narrower from peripheral to central AuAu collisions 
and acquiring more $\eta$-$\phi$ symmetry. 
We hypothesize that this reflects a change in the hadronization geometry from one dominated
by independent string break-ups in peripheral collisions
to a  bulk hadronization process in which individual strings are
no longer relevant.


\section{Charge-independent number correlations -- longitudinal minijet deformation}
\label{Sec:CI}
The charge-independent (CI) number correlation sums the like-sign and the unlike-sign correlations.
In AuAu collisions, these correlations have been analyzed by STAR
in the  space of pseudorapidity $\eta$ and azimuthal angle $\phi$~\cite{STAR_wavelet,STAR_CI_axial}
and in the transverse momentum space~\cite{STAR_CI_mtmt}.

Our discrete-wavelet technique~\cite{STAR_wavelet} extracts information about correlation
structure by measuring the power spectra of local fluctuations in the density
of charged hadrons with respect to a mixed-event reference (the so-called "dynamic texture"
of the event).
As discussed in~\cite{QM2004_poster}, this observable 
is a measure of the gradient
of the two-particle correlation function, but takes less computing.
\footnote{
The number of computations required is $\mathcal{O}(N)$, in contrast with
 $\mathcal{O}(N^2)$ required for direct construction of the correlation function.
}

The experiment ~\cite{STAR_wavelet} reveals a reduction in the dynamic texture
(and  thus in  the correlation function gradient) along the $\eta$-direction at $p_t>0.6$ GeV/$c$
in the central AuAu collisions, compared to an expectation based on peripheral events.
HIJING simulations~\cite{STAR_wavelet}
point to minijets as the likely source of correlations of this scale
and magnitude. 
The possible mechanism of the reduction is a coupling between the minijet fragments 
and the longitudinally expanding bulk medium.
In paper ~\cite{STAR_CI_axial}, the same correlation structure
is investigated in the joint $\eta$-$\phi$ 
autocorrelation technique.
It is found that the correlations are broadened in $\eta_\Delta$ with centrality.
A reduction in the  correlation gradient ("dynamic texture") and the elongation
of the correlation function along $\eta_\Delta$ are consistent descriptions of the effect.


\section{$p_t$ correlations --minijets and elliptic flow}
\label{Sec:pt_correlation}
\begin{figure}
\begin{minipage}[t]{9pc}
\includegraphics[width=9pc]{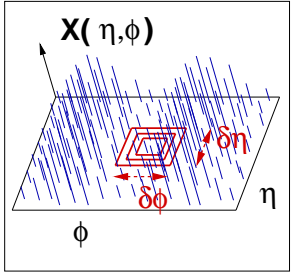}
\end{minipage}
\hspace{2pc}%
\begin{minipage}[b]{27pc}
\caption{
\label{Fig:scaled_analysis}
A fantasy event.  Positions of spikes indicate $\eta,\phi$ kinematics
of individual particles, their heights -- magnitude of quantity $X$
assigned to a particle, which could be $p_t$ or a derived quantity.
 This event has azimuthal
anisotropy, however its $v_2$ may be 0, illustrating the difference
between number and $p_t$ flow.  The analysis 
proceeds by selecting a bin of size $\delta\eta,\delta\phi$,
integrating the {\it random field} $X$ within the bin, finding the
variance of that integral in an event sample, then feeding this
information into an integral equation to yield the correlation
of $X$, as discussed in \ref{Sec:integral_equation}.} 
\end{minipage}
\end{figure}

We require a direct measurement of correlations in the velocity field,
independent of particle number correlations.
To form an independent measurement, these have to be separated from
particle number correlations.  For a random field (such as particle
number or velocity distributed in $\eta$ and $\phi$, see Fig.\ref{Fig:scaled_analysis})
variance of the integrated content of a bin can be related via an integral
equation to the correlation function of the
field~\cite{van_Marcke_text,Trainor_inversion}.  The correlation
function can be reconstructed from the measured variance by solving the
integral equation derived in~\cite{van_Marcke_text,Trainor_inversion}.  
This process
offers a computational advantage since the bin integration involves
$\mathcal{O}(N)$ computations, whereas two-particle correlations
require $\mathcal{O}(N^2)$ ($N$ is the event multiplicity used).  
By
varying the bin size one varies the range of the difference variable in
the correlation function.

\begin{figure}
\begin{minipage}[t]{52mm}
\includegraphics[width=52mm]{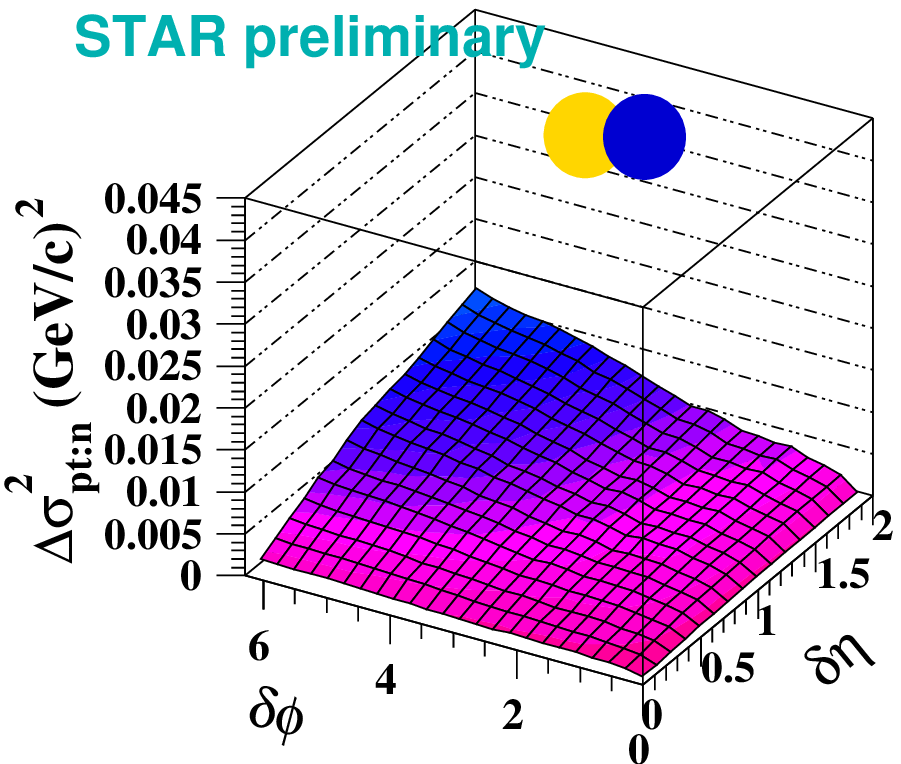}
\includegraphics[width=52mm]{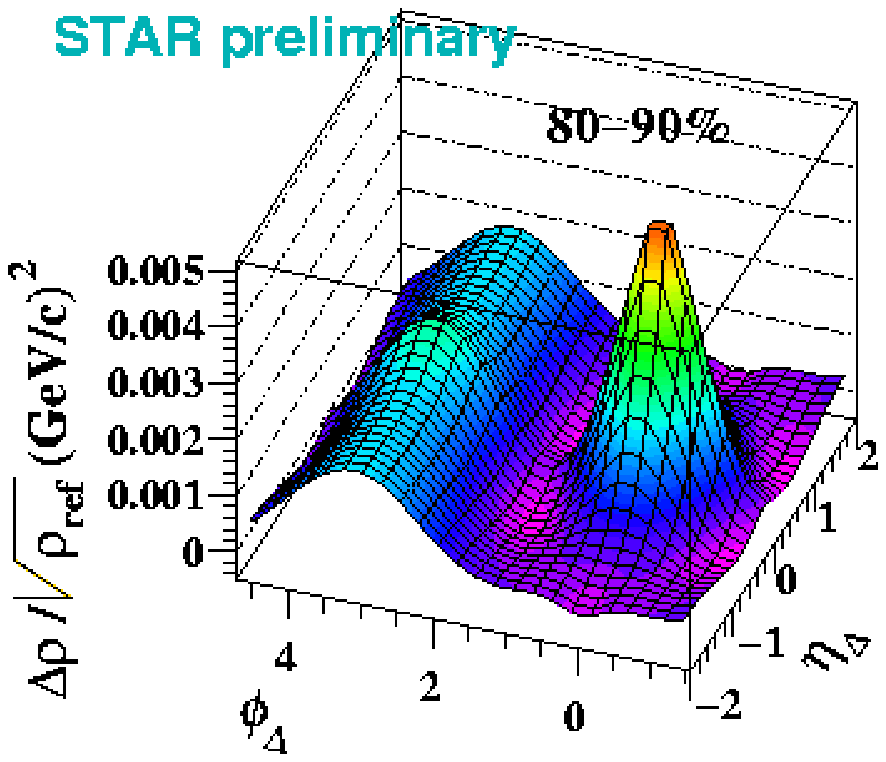}
\end{minipage}
\begin{minipage}[t]{52mm}
\includegraphics[width=52mm]{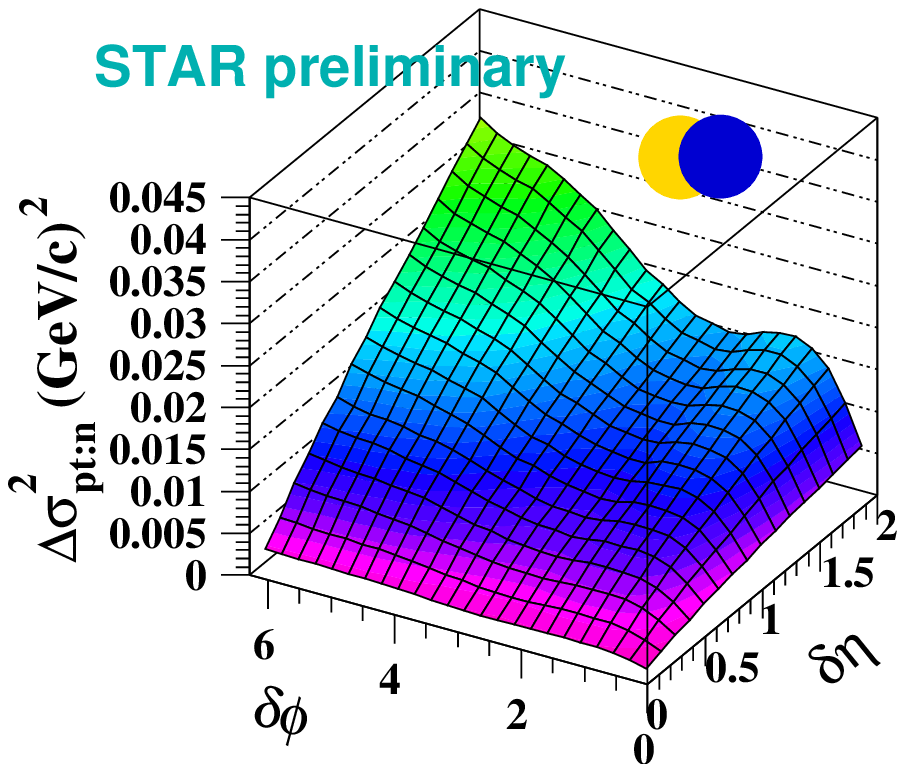}
\includegraphics[width=52mm]{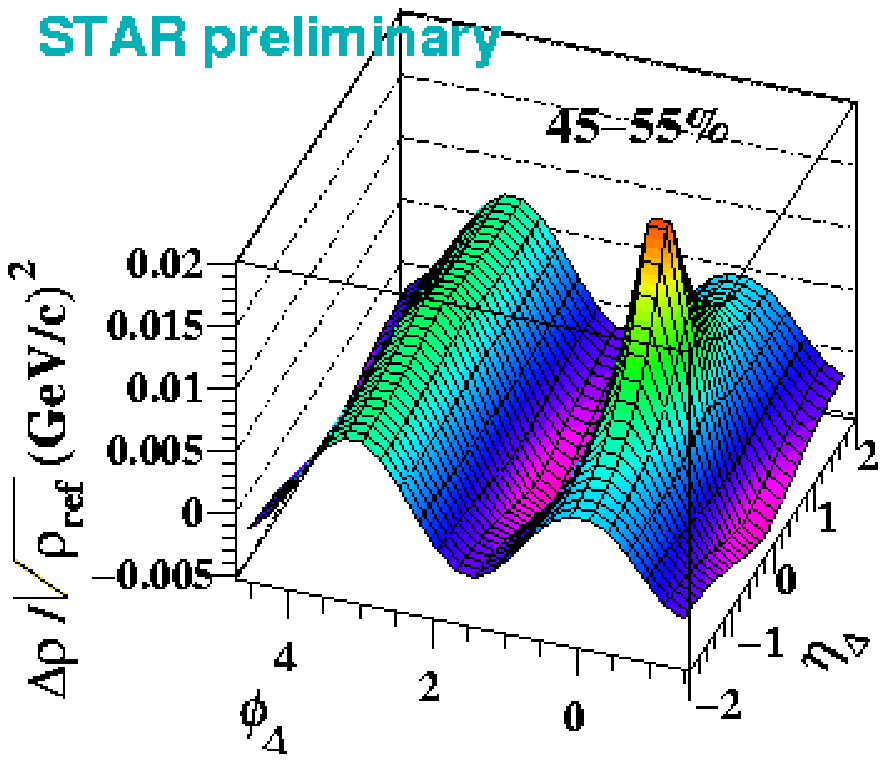}
\end{minipage}
\begin{minipage}[t]{52mm}
\includegraphics[width=52mm]{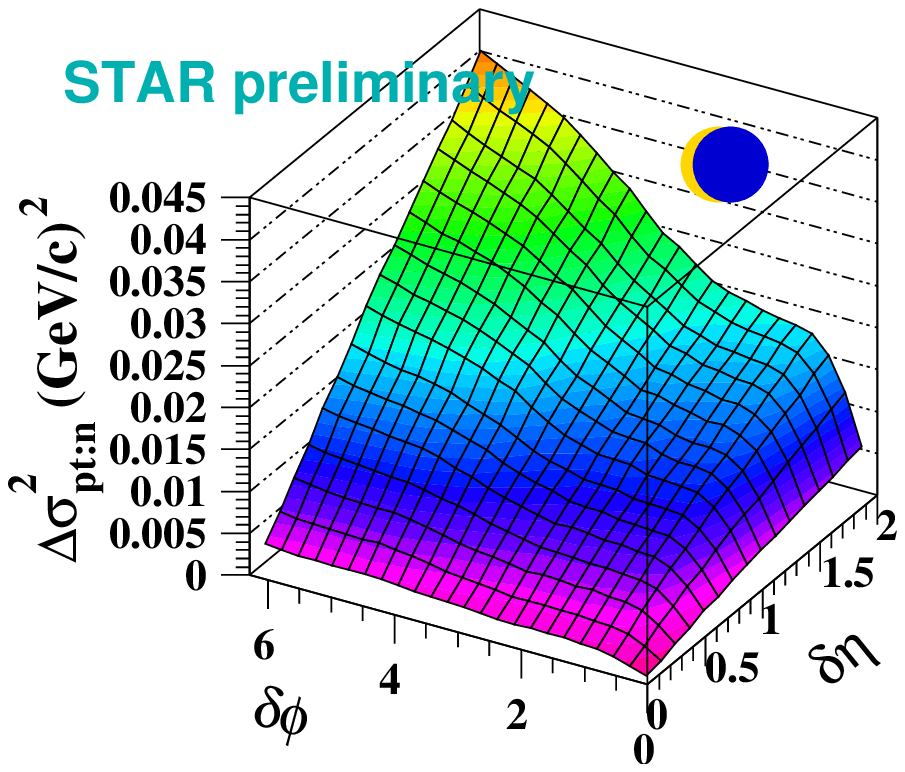}
\includegraphics[width=52mm]{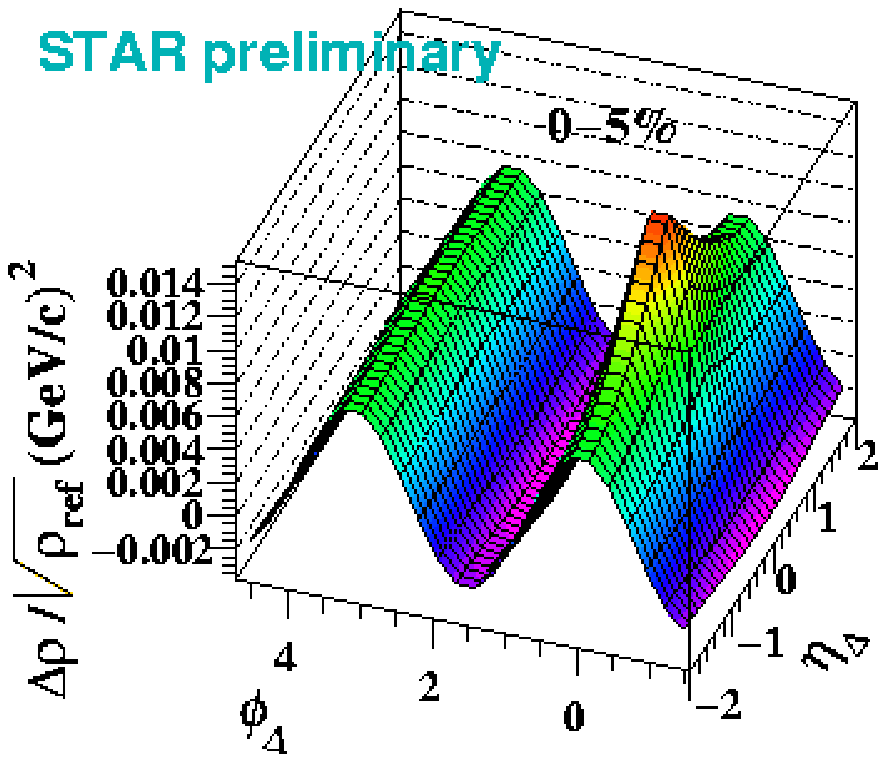}
\end{minipage}
\caption{\label{Fig:pt_n_auto}
Top: scale dependence of the reference-subtracted variance $\Delta
\sigma^2_{p_t:n}$.
 Bottom: reference-subtracted normalized
correlation functions reconstructed from $\Delta \sigma^2_{p_t:n}$
by inversion. The percentage intervals indicate centrality for each column-wise
pair of plots.}
\end{figure}
\begin{figure}
\includegraphics[width=52mm]{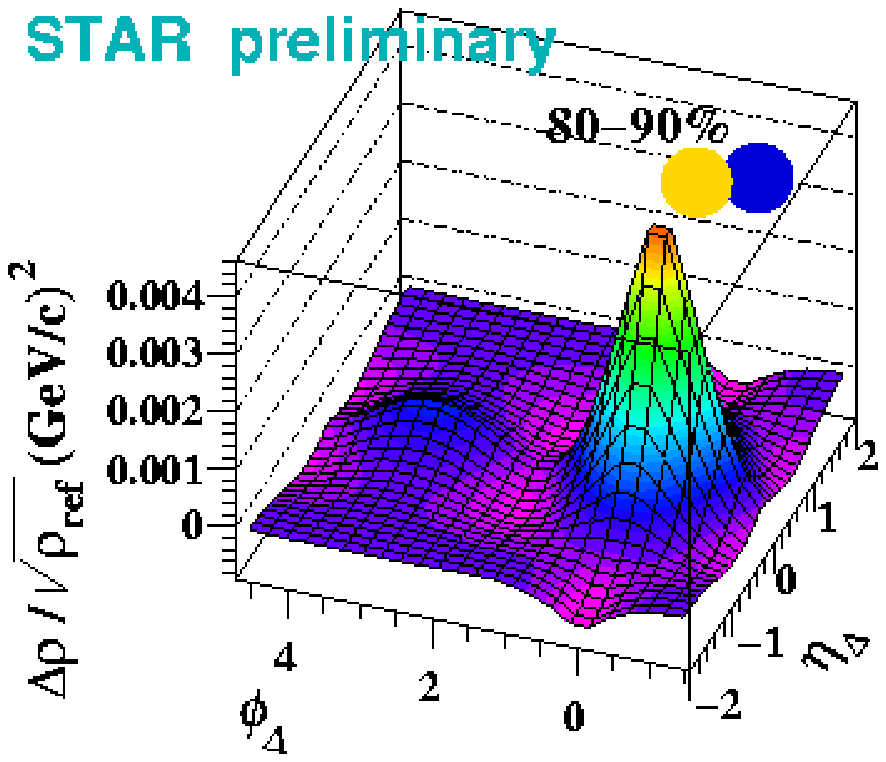}
\includegraphics[width=52mm]{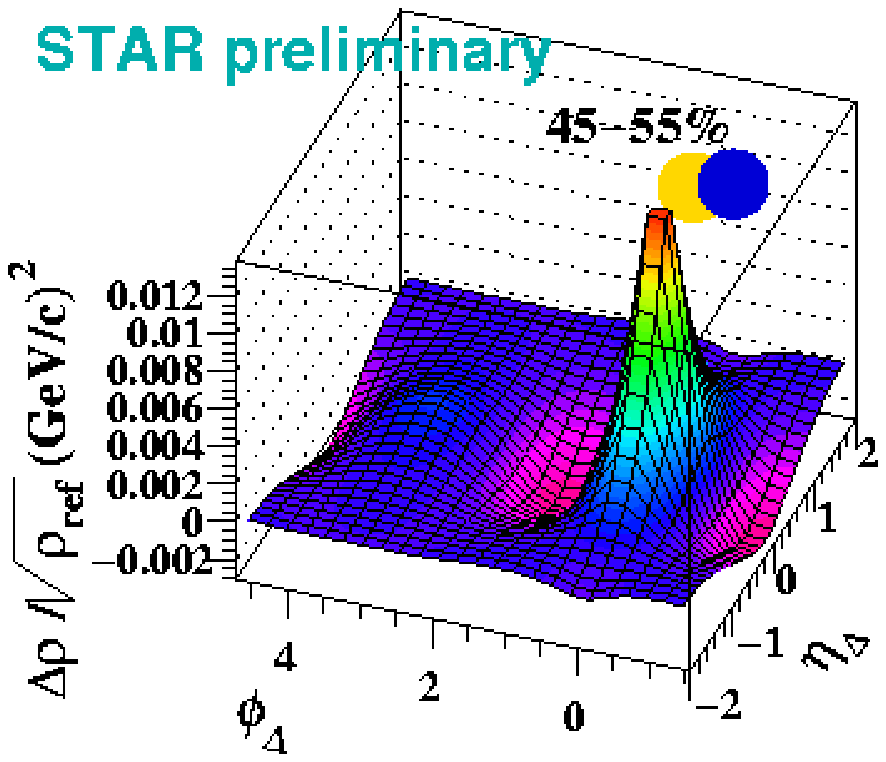}
\includegraphics[width=52mm]{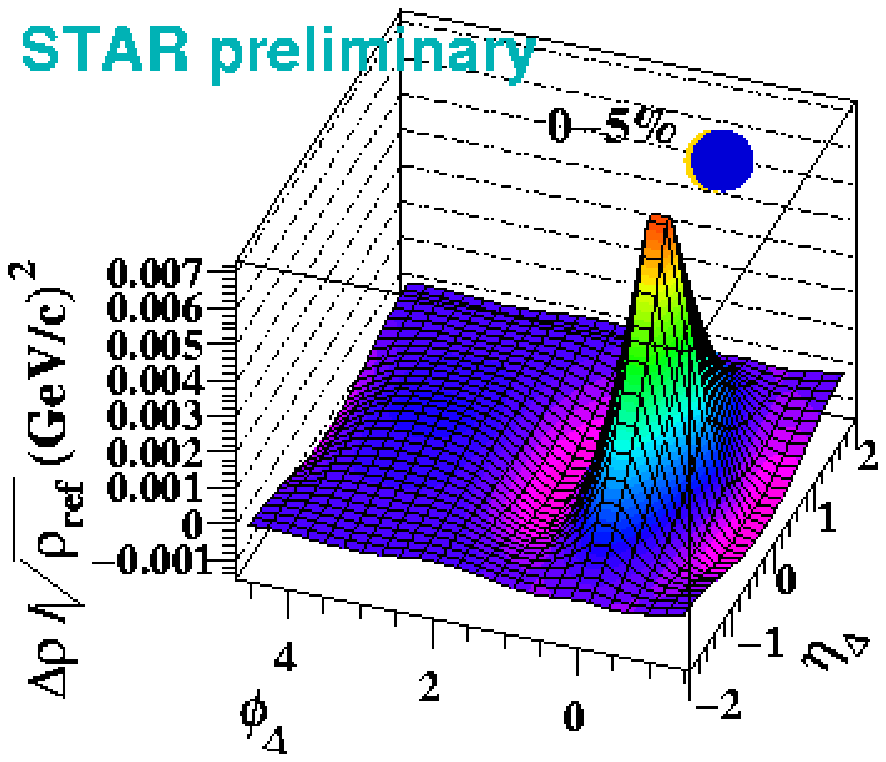}
\caption{\label{Fig:pt_n_no_v}
Same data as in Fig.\ref{Fig:pt_n_auto} with subtracted azimuthal harmonics
$v_1$ and $v_2$.
}
\end{figure}

In order to extract the $p_t$ correlation alone, we
need to disentangle the effect of $n$ from that of  $p_{t,i}$ in the variance of
$p_t \equiv \sum_{i \in (\eta,\phi) \rm{bin}} p_{t,i}$, and from statistical fluctuations.
We denote the desired variance  as $\Delta\sigma^2_{p_t:n}$.
As an estimator of  $\Delta\sigma^2_{p_t:n}$, we use 
${\rm Var}[p_t-n\hat{p_t}]/\bar{n}-\sigma^2_{\hat{p_t}}$,
where the latter term is the inclusive per-hadron $p_t$ variance.
In the case when $p_t$ in the bin, event-by-event, derives from the same 
parent distribution (not the case in Fig.\ref{Fig:scaled_analysis}), $\Delta\sigma^2_{p_t:n}$
converges to 0, despite fluctuations (statistical or not) in $n$.

Top panels in Fig.\ref{Fig:pt_n_auto} represent scale dependence of the variance 
$\Delta\sigma^2_{p_t:n}$
for three centrality classes.
The bottom panels, representing
per-particle normalized correlations plotted on difference variables,
 are obtained from the respective top panels
by solving the integral equation derived in ~\cite{van_Marcke_text,Trainor_inversion}
(see \ref{Sec:integral_equation})
which relates variance and correlation.

While the quadrupole wave in Fig.\ref{Fig:pt_n_auto}(bottom row)
is the reflection of the elliptic flow in the $p_t:n$ correlation function,
the peak at 0 angle differences likely reflects minijets~\cite{Trainor_HIJING_minijets}.
But unlike flow, the effect of minijets on the HBT radii at RHIC energy
is not  part of the standard (local equilibrium) picture.
Therefore this effect may constitute an under-appreciated factor driving the effective
radii (Fig.\ref{Fig:HBT_sqrt_s}) down at higher $\sqrt{s_{NN}}$.

The amplitude of the minijet peak has a  non-monotonic behavior with increasing centrality, 
rising in peripheral and  falling in central collisions.
The peak is seen to be elongated longitudinally in central collisions, showing qualitatively
the same behavior as seen with number correlations by using wavelet analysis and 
joint autocorrelation (Section\ref{Sec:CI}).

\section{Conclusions}

The small effective HBT radii at RHIC indicate that the system is
highly inhomogeneous and decouples rapidly.  STAR has evidence that
semi-hard scattering leaves a trace in the "soft" $p_t$ domain which
is usually considered a safe realm for local equilibrium
approximations, thus possibly explaining some of the inhomogeneity.
  Whether or not the minijet correlations present a
challenge to the picture of local equilibrium which
constitutes a basis for much of the HBT interpretation infrastructure,
will depend on the effective final-state-particle number involved in
a minijet, or in other words, on the {\it order} of those correlations
(higher order being more hydro-like and thus probably less of a
challenge).  Therefore a study of higher order correlations and
cumulants may be a promising future direction.
The phenomenon of $\eta$-broadening of minijets at ``soft''
 $p_t$ (Section \ref{Sec:CI} and \ref{Sec:pt_correlation})
can clarify the nature of the medium created in
central AuAu collisions, provided that the coupling mechanism, apparently sensitive to that 
nature, is under theoretical control.


\appendix
\section{Notation}
\label{Sec:Notation}
In our notation
$i$ is particle index,
$n$  is number of particles within a kinematic cut (bin),
$\overline{(\dots)}$ denotes an average over events,
$\hat{p_t}$ is an inclusive mean $p_t$ per particle,
$x_\Delta$ (variants:  $\eta_\Delta$, $\phi_\Delta$) is difference
variable $=x_i-x_{i'}$,
$\delta x$  is scale (range of local integration, see Fig.\ref{Fig:scaled_analysis}),
$\Delta x$ is the upper limit on $\delta x$.
Covariance of $x$ and $y$ is denoted as ${\rm Cov}[x,y]$,
variance of $x$ -- as ${\rm Var[x]}$.

\section{From variance to correlation}
\label{Sec:integral_equation}
The connection between variance and correlation function -- the basis of analysis
in Section\ref{Sec:pt_correlation} -- is discussed in detail using integral calculus in~\cite{van_Marcke_text}
and the algebra of discrete bins in ~\cite{Trainor_inversion}.
For a random field $X(t)$, the {\it autocorrelation} function is defined~\cite{Flandrin} as
\begin{equation}
\rho(X,t_\Delta) \equiv \overline{X(t)X(t+t_\Delta)}
\end{equation}
(assuming that correlations do not depend on position $t$).

The variance of  $X$, integrated over bin of width $\delta\eta,\delta\phi$, centered at 0
(see Fig.\ref{Fig:scaled_analysis}) is
\begin{eqnarray}
{\rm Var}[X;\delta\eta,\delta\phi] =
\overline{\left(\int_{-\delta\eta/2}^{\delta\eta/2}\int_{-\delta\phi/2}^{\delta\phi/2} X \,d\eta \,d\phi\right)^2}
-
\left(\int_{-\delta\eta/2}^{\delta\eta/2}\int_{-\delta\phi/2}^{\delta\phi/2} \overline{X} \,d\eta \,d\phi \right)^2
=\\
\int_{-\delta\eta/2}^{\delta\eta/2} \,d\eta_1   \int_{-\delta\phi/2}^{\delta\phi/2} \,d\phi_1
\int_{-\delta\eta/2}^{\delta\eta/2} \,d\eta_2   \int_{-\delta\phi/2}^{\delta\phi/2} \,d\phi_2 
[\overline{X(\eta_1,\phi_1)X(\eta_2,\phi_2)}-
\overline{X(\eta_1,\phi_1)}\times
\overline{X(\eta_2,\phi_2)}]
\end{eqnarray}
As we form the reference-subtracted variance
$\Delta\sigma^2_X \equiv {\rm Var}[X;\delta\eta,\delta\phi] - 
{\rm Var}[X_{\rm ref};\delta\eta,\delta\phi]$,
 the first term in the  brackets above 
becomes the reference-subtracted
correlation 
$\Delta\rho(X,\eta_\Delta,\phi_\Delta)$
and the second term cancels with the reference since $\overline{X} = \overline{X_{\rm ref}}$.
\begin{eqnarray}
\Delta\sigma^2_X  =
\int_{-\delta\eta/2}^{\delta\eta/2} \,d\eta_1   \int_{-\delta\phi/2}^{\delta\phi/2} \,d\phi_1
\int_{-\delta\eta/2}^{\delta\eta/2} \,d\eta_2   \int_{-\delta\phi/2}^{\delta\phi/2} \,d\phi_2
\Delta \rho(X,\eta_1-\eta_2,\phi_1-\phi_2)
\\ 
= 2 \int_{0}^{\delta\eta}\,d\eta_\Delta 2 \int_{0}^{\delta\phi}\,d\phi_\Delta
(\delta\eta - \eta_\Delta)
(\delta\phi - \phi_\Delta)
\Delta
\rho(X,\eta_\Delta,\phi_\Delta)
\label{Eq:variable_change}
\end{eqnarray}
Eq.\ref{Eq:variable_change} is obtained by making a change of the integration variables
from $\eta_1$,$\eta_2$ to $\eta_1$ and $\eta_\Delta=\eta_1-\eta_2$, integrating over $\eta_1$,
and repeating the same for $\phi$ (see ~\cite{van_Marcke_text} for details).
In the actual analysis, this integral equation is discretized by partitioning the acceptance 
into  $m_\delta\times n_\delta$ {\it microbins} of size $\varepsilon_\eta\times\varepsilon_\phi$.
$X$ is replaced by $(p_t-n\hat{p_t})/\sqrt{\bar{n}}$.
This denominator creates the reference number 
correlation $\rho_{\rm ref}(n)$ in the final expression \ref{Eq:final_discretized}
which makes it a per-particle measure:
 
\begin{equation}
\Delta\sigma^2_{p_t:n}(m_\delta \varepsilon_\eta,n_\delta \varepsilon_\phi)
= 4 \sum_{k,l=1}^{m_\delta,n_\delta}\varepsilon_\eta \varepsilon_\phi K_{m_\delta n_\delta:kl}
\frac{\Delta \rho(p_t:n;k\varepsilon_\eta,l\varepsilon_\phi)}
{\sqrt{\rho_{\rm ref}(n;k\varepsilon_\eta,l\varepsilon_\phi)}}
\label{Eq:final_discretized}
\end{equation}

Here the kernel $K$ is a discrete representation of the continuous case :
\begin{equation}
(\delta\eta - \eta_\Delta)(\delta\phi - \phi_\Delta) \rightarrow
\varepsilon_\eta\varepsilon_\phi
K_{m_\delta n_\delta:kl} \equiv
\varepsilon_\eta\varepsilon_\phi
(m_\delta - k +\frac{1}{2})(n_\delta - l +\frac{1}{2})
\end{equation}

Knowing the computationally cheaper quantity $\Delta\sigma^2_{p_t:n}$, 
we solve the integral equation \ref{Eq:final_discretized}
for $\Delta \rho/\sqrt{\rho_{\rm ref}}$ using standard numerical techniques~\cite{Trainor_inversion}.

\end{document}